\documentclass[aps,prl,groupedaddress,reprint]{revtex4-1}
\usepackage{graphicx}
\usepackage{ulem}
\usepackage{amssymb}
\usepackage{amsmath}
\usepackage{todonotes}
\begin{document}
%\draft                    

%\flushbottom
%\twocolumn[
%\hsize\textwidth\columnwidth\hsize\csname @twocolumnfalse\endcsname

\title{A universal formula for the field enhancement factor}
 
%\vskip 0.3 in

\author{Debabrata Biswas}
%\author{\ldots}

\affiliation{
Bhabha Atomic Research Centre,
Mumbai 400 085, INDIA \\
Homi Bhabha National Institute, Mumbai 400 094}

%\pacs{85.45.-w}{}
%\pacs{03.65.Sq}{}
%\pacs{03.65.Xp}{}
%\pacs{52.59.Sa}{}

\begin{abstract}
  The field enhancement factor (FEF) is an important quantity
  in field emission calculations since the tunneling electron current depends very
  sensitively on its magnitude. The exact
  dependence of FEF on the emitter height $h$, the radius of curvature at the apex $R_a$,
  as well as the shape of the emitter base is still largely unknown. In this work,
  a universal formula for the field enhancement factor is derived.
  It depends on the ratio $h/R_a$ and has the form
  $\gamma_a = (2h/R_a)/[\alpha_1 \log(4h/R_a) - \alpha_2 ]$ where
  $\alpha_1$, $\alpha_2$ depend on the charge distribution on the emitter.
  Numerical results show that a simpler form $\gamma_a = (2h/R_a)/[\log(4h/R_a) - \alpha]$
  is equally valid with $\alpha$ depending on the class of emitter and indicative of
  the shielding by the emitter-base. For the hyperboloid, conical and  ellipsoid
  emitters, the value of $\alpha$ is  $0, 0.88$ and $2$ while for the cylindrical base where
  shielding is minimum, $\alpha \simeq 2.6$. 
\end{abstract}

%\email{dbiswas@barc.gov.in}
%\maketitle

%\pacs{85.45.-w}{}
%\pacs{03.65.Sq}{}
%\pacs{03.65.Xp}{}
%\pacs{52.59.Sa}{}

%\date{\today}
%\vskip 0.2 in
%\centerline{\bf Abstract}

%\vskip 0.25 in

%\pacs{85.35.-p, 03.65.Sq, 52.59.Sa}

\maketitle

%]
\newcommand{\be}{\begin{equation}}
\newcommand{\ee}{\end{equation}}
\newcommand{\bea}{\begin{eqnarray}}
\newcommand{\eea}{\end{eqnarray}}
\newcommand{\Tbar}{{\bar{T}}}
\newcommand{\En}{{\cal E}}
\newcommand{\K}{{\cal K}}
\newcommand{\GC}{{\cal \tt G}}
\newcommand{\Lop}{{\cal L}}
\newcommand{\DB}[1]{\marginpar{\footnotesize DB: #1}}
\newcommand{\q}{\vec{q}}
\newcommand{\kt}{\tilde{k}}
\newcommand{\Lopn}{\tilde{\Lop}}
\newcommand{\noi}{\noindent}
\newcommand{\ovn}{\bar{n}}
\newcommand{\ovx}{\bar{x}}
\newcommand{\ovE}{\bar{E}}
\newcommand{\ovV}{\bar{V}}
\newcommand{\ovU}{\bar{U}}
\newcommand{\ovJ}{\bar{J}}
\newcommand{\calE}{{\cal E}}
\newcommand{\ovphi}{\bar{\phi}}
\newcommand{\zt}{\tilde{z}}
\newcommand{\rt}{\tilde{\rho}}
\newcommand{\tth}{\tilde{\theta}}
\newcommand{\nuv}{{\rm v}}
\newcommand{\ck}{{\cal K}}
\newcommand{\cc}{{\cal C}}
\newcommand{\ca}{{\cal A}}
\newcommand{\cb}{{\cal B}}
\newcommand{\cg}{{\cal G}}
\newcommand{\ce}{{\cal E}}
\newcommand{\fn}{{\small {\rm  FN}}}
\newcommand\norm[1]{\left\lVert#1\right\rVert}

%\newpage
%\noindent

%\section{Introduction}
%\label{sec:Introduction}

\section{Introduction}
\label{sec:intro}

The field enhancement factor (FEF) is the key to understanding the phenomenon
of field emission \cite{FN,murphy,forbes,forbes_deane,fursey,jensen_ency,liang,egorov_springer,db2018a}.
In the neighbourhood of a sharp emitter tip, an external electric field gets
enhanced, sometimes by several orders of magnitude. This can lead to electron emission
even when the external field is of moderate strength. The high local field
near the apex helps in reducing the height and width of the potential barrier making it
easier for electrons to tunnel through. The phenomenon is useful where a source of cold
electrons is required (such as in vacuum microwave and terahertz devices, microscopy,
lithography and space and medical applications) but is undesirable in devices
where a bad surface finish can lead to vacuum breakdown \cite{fursey}.

Over the past decades, several models have been proposed to understand field enhancement.
It is thus known that FEF is a geometric quantity that is independent of the
magnitude of the external field. Rather, it seems to depend on the ratio
$h/R_a$ where $h$ is the height of the emitter and $R_a$ is the apex radius of curvature.
Analytical results for the field enhancement factor, $\gamma_a$, are known for the ellipsoid and hyperboloid emitters
and can be expressed (for large $h/R_a$) as

\be
\gamma_a = \frac{2h/R_a}{\log(4h/Ra) - \alpha}  \label{eq:fef}
\ee

\noi
with $\alpha = 2$ for the ellipsoid and $\alpha = 0$ for the hyperboloid.
Other shapes that are analytically not solvable, have been studied in
several ways \cite{forbes2003,podenok,read,kokkorakis,bonard,zhbanov,forbes2016}.
The ``floating-sphere at emitter plane potential'' is a model
for emitter tips that over-estimates the enhancement factor but nonetheless
is considered useful in qualitative predictions. Its simplest prediction for a single emitter 
is $\gamma_a = h/R_a + 3.5 + \mathcal{O}(R_a/h)$. Numerically, carbon nanotubes have been modelled as hemisphere on a cylindrical
post. The results 
suggest fits of the form $\gamma_a \simeq 0.7 (h/R_a)$  while more
elaborate ones 
are expressed as $\gamma_a \simeq a(b + h/R_a)^\sigma$ with $0.9 < \sigma \leq 1$.
These expressions for $\gamma_a$ seem to be very different from the ellipsoid and
hyperboloid results and can at best be local approximations of a more general
expression. A formula, applicable to a wide class of emitters has thus been elusive
and it is our aim here to provide one that is universally applicable.

We shall approach the problem analytically using a general nonlinear line charge distribution and provide
a formula that is universally valid and reduces to  Eq.~\ref{eq:fef} for the ellipsoid.
However, it depends on properties of the charge distribution that are {\it a priori} unknown.
We therefore address the problem numerically and establish that Eq.~1 is still valid
and $\alpha$ is an indicator of the absence of shielding by the emitter base.
Thus, for the hyperboloid, $\alpha = 0$ while for an emitter
top placed on a cylindrical post, $\alpha \simeq 2.6$, indicative
of the absence of shielding.

\section{The Line Charge Model}

The problem of an emitter of height $h$ placed on grounded metallic plane and aligned along
an external electrostatic field $-E_0 \hat{z}$, can be modelled by a vertical line charge
distribution and its image. Denoting the line charge density by $\Lambda(s)$, the potential
at any point ($\rho,z$) can be expressed as \cite{jap2016,db_ultram}

\be
\begin{split}
V(\rho,z) = & \frac{1}{4\pi\epsilon_0}\Big[ \int_0^L \frac{\Lambda(s)}{\big[\rho^2 + (z - s)^2\big]^{1/2}} ds ~
  - \\
  &  \int_0^L \frac{\Lambda(s)}{\big[\rho^2 + (z + s)^2\big]^{1/2}} ds \Big] + E_0 z \label{eq:pot}
\end{split}
\ee

\noi
where $L$ is the extent of the line charge distribution and $E_0$ is the magnitude of
the electric field. The zero-potential contour then
corresponds to the surface of the emitter and quantities such as the enhancement factor
and the principle radii of curvature can be calculated numerically. The line charge
distribution can be thought of as a projection of the surface charge
of the emitter along the axis.

Amongst the unknowns in Eq.~\ref{eq:pot} are the parameters defining the line charge distribution
and its extent $L$. These can in principle be calculated by imposing the requirement that
the potential should vanish along the surface of the emitter. A shortcoming of the line charge model
is the absence of image charges due to the anode in the formulation of the problem. The results
are therefore valid when the anode is sufficiently far away from the emitter. In the rest of the manuscript,
we shall consider axially symmetric emitters and assume $\Lambda(s)$ to be a nonlinear
function of $s$ unless stated otherwise.

\section{The Apex Field Enhancement Factor}

We are interested in the field enhancement factor, $\gamma_a$ at the emitter apex. For axially
symmetric emitters aligned along $\hat{z}$, this is defined as
$\gamma_a = - \frac{1}{E_0} \frac{\partial V}{\partial z} {|_{\rho=0,z=h}}$.
Note that for parabolic emitter tips, it has recently been established \cite{db_ultram} that $\gamma = \gamma_a \cos\tth$
where $\gamma$ is the enhancement factor at a point ($\rho,z$) on the emitter surface while
$\cos\tth = (z/h)/[ (\rho/R_a)^2 + (z/h)^2 ]^{1/2}$.
Thus, the local field around the emitter apex can be determined if the apex field
enhancement factor is known.

Our starting point for the FEF is Eq.~\ref{eq:pot}. At the apex,

\be
\begin{split}
  \frac{\partial V}{\partial z} {|_{(\rho=0,z=h)}} = & - \frac{1}{4\pi\epsilon_0}\Big[\int_0^L \frac{\Lambda(s)}{(h-s)^2} ds~ - \\
    &  \int_0^L \frac{\Lambda(s)}{(h+s)^2} ds   \Big] + E_0
\end{split}
\ee

\noi
so that on writing $\Lambda(s) = s f(s)$, we have

\be
\begin{split}
  \frac{\partial V}{\partial z} {|_{(\rho=0,z=h)}} = &  - \frac{1}{4\pi\epsilon_0} \Big[ \int_0^L ds \Big\{ -\frac{f(s)}{h-s}
    +  \\
    & \frac{h f(s)} {(h-s)^2} -   \frac{f(s)}{h+s} + \frac{h f(s)}{(h + s)^2} \Big\} \Big] + E_0.
\end{split}
\ee

\noi
Using partial integrations,

\be
\begin{split}
  \frac{\partial V}{\partial z} {|_{(\rho=0,z=h)}} = &   \frac{1}{4\pi\epsilon_0} \Big[ f(L) \ln\Big(\frac{h+L}{h-L)}\Big) (1 - C_1) \\
    & - f(L) \frac{2hL}{h^2-L^2}( 1 - \cc_0) \Big] + E_0  \label{eq:Vz}
\end{split}
\ee

\noi
where

\bea
\cc_0 & = & \int_0^L \frac{f'(s)}{f(L)} \frac{s/(h^2 - s^2)}{L/(h^2 - L^2)}  ds \label{eq:C0} \\
\cc_1 & = & \int_0^L \frac{f'(s)}{f(L)}\frac{\ln\Big(\frac{h+s}{h-s}\Big)}{\ln\Big(\frac{h+L}{h-L}\Big)} ds.
\label{eq:C1}
\eea

\noi
Note that $L$ is the height of the line charge distribution and must extend almost till the apex height $h$.
Moreover, since the charge distribution is well behaved and can be expressed as a polynomial function of
degree $n$ (for cases of interest here, $n \leq 5$), it obeys Bernstein's inequality \cite{ineq}

\be
|f'(x)| \leq \frac{n}{(1 - x^2)^{1/2}} \norm{f}  \label{eq:bern}
\ee

\noi
where $x \in [-1,1]$ and $\norm{f}$ denotes the maximum value of $f$ in this interval. With
$x = s/h$ and applying the inequality, it can be shown that  $\cc_0 \sim (h^2 - L^2)^{1/2}$ is
vanishingly small for sharp-tipped emitters. In contrast, $\cc_1$ cannot be neglected due to the logarithmic dependence.

\subsection{Linear line charge density - the ellipsoid}

For a linear line charge as in case of an ellipsoidal emitter, $f'(s) = 0$ so that $\cc_0 = 0 = \cc_1$.
Denoting $f(s) = \lambda$, a constant, the enhancement factor can be expressed
as

\bea
\gamma_a & = & \frac{|\lambda|}{4\pi\epsilon_0 E_0}  \Big[  
    \frac{2hL}{h^2-L^2} - \ln\Big(\frac{h+L}{h-L)}\Big) \Big] - 1 \\
& \simeq & \frac{|\lambda|}{4\pi\epsilon_0 E_0}  \Big[ 
    \frac{2h^2}{h^2-L^2} - \ln\Big(\frac{4h^2}{h^2 - L^2}\Big) \Big] - 1
\eea

\noi
where the last line assumes that $h \simeq L$ as the line charge must extend almost till the apex height $h$.
Further, if $h^2/(h^2 - L^2)$ is large, the logarithmic part can be neglected so that

\be
\gamma_a = \frac{|\lambda|}{4\pi\epsilon_0 E_0}  \Big[ \frac{2h^2}{h^2-L^2}  \Big]
\ee

\noi
The parameter $\lambda$ can be determined by demanding that the potential in Eq.~\ref{eq:pot} vanishes
at the apex for $\Lambda(s) = \lambda s$. Thus,

\be
\lambda = - \frac{4\pi\epsilon_0 E_0 h}{h\ln\Big(\frac{h+L}{h-L}\Big) - 2L}
\ee

\noi
using which, the field enhancement factor under the approximation $L \simeq h$ can be
expressed as

\be
\gamma_a = \frac{2h/R_0}{\ln(4h/R_0) - 2}  \label{eq:ellipsoid}
\ee  

\noi
where $R_0 = (h^2 - L^2)/h$. We show in the appendix that $R_0 \simeq R_a$, the apex radius curvature for
general sharp emitters..
We are thus able to recover the field enhancement factor for an ellipsoid at least when $h/R_a$ is
large.

\subsection{The nonlinear density}

The general case corresponding to nonlinear $\Lambda(s)$ holds greater interest. Since
$h \simeq L$, under approximations mentioned earlier, the enhancement factor reduces to

\be
\gamma_a = \frac{|f(L)|}{4\pi\epsilon_0 E_0}  \frac{2h^2}{h^2-L^2} \label{eq:nonlin}
\ee

\noi
where $f(L)$ can be determined following the procedure for $\lambda$.

As in the linear case, $\Lambda(L)$ can be determined by demanding that the potential
vanishes at the emitter apex, ($\rho = 0, z = h$). Thus,

\bea
4\pi & \epsilon_0 &  E_0 h  =   \int_0^L ds \Big[ \frac{s f(s)}{h+s} - \frac{s f(s)}{h-s} \Big] \\
& = & \int_0^L ds \Big[ f(s) - \frac{h f(s)}{h+s} + f(s) - \frac{h f(s)}{h-s} \Big].
\eea

\noi
On integrating by parts,

\be
\begin{split}
  4\pi \epsilon_0  E_0 h  = - &  \Big[ f(L) h \ln\Big(\frac{h+L}{h-L} \Big)(1 - \cc_1) - \\
    & f(L)2L(1 - \cc_2) \Big]  
\end{split}
  \ee

  \noi
  where $\cc_2  =  \int_0^L \frac{f'(s)}{f(L)} \frac{s}{L}  ds$. Thus,
%  \cc_2 = \int_0^L ds \Big[ \big\{2s - h \ln\big(\frac{h+s}{h-s} \Big)\big\} \Big] \frac{f'(s)}{f(L)}.

\be
  f(L) = - \frac{4\pi \epsilon_0  E_0 h}{\Big[ h \ln\Big(\frac{h+L}{h-L} \Big)(1 - \cc_1) - 2L(1 - \cc_2) \Big]}
\ee

\noi
and hence with $L \simeq h$,

\be
\gamma_a  \simeq  \Big[  \frac{2h/R_0}{\alpha_1 \ln\big(4h/R_0\big) - \alpha_2 }\Big] \label{eq:main}
\ee

\begin{figure}[htb]
%\vskip -2.1 cm
%\hskip -1.0cm
%\centering
\hspace*{-1.0cm}\includegraphics[width=0.5\textwidth]{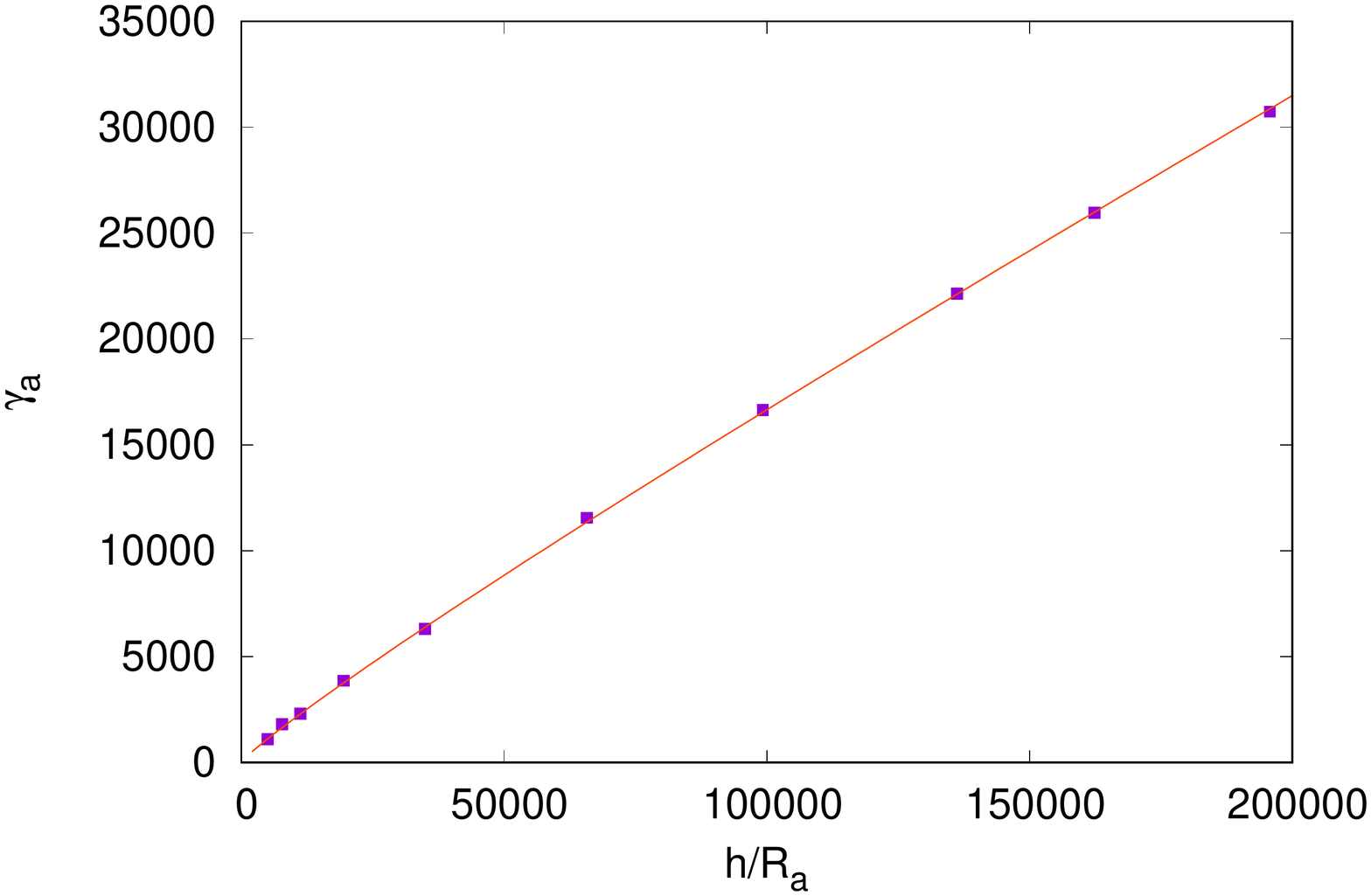}
\vskip -0.9 cm
\caption{The field enhancement factor for parabolic emitter tops with a conical base (squares)
  with the best fit using Eq.~\ref{eq:fef} (continuous line).
}
\label{fig:angular}
\end{figure}

\noi
where $\alpha_1 = 1 - \cc_1$, $\alpha_2 = 2(1 - \cc_2)$ and $R_0 = (h^2 - L^2)/h \simeq R_a$.
Eq.~\ref{eq:main} is thus an expression for the field enhancement
factor that is generally applicable to all shapes. The correction terms $\cc_1$ and $\cc_2$
are however {\it a priori} unknown since they depend on details of the charge
distribution. We have tested Eq.~\ref{eq:main} numerically for several emitter shapes. For
example, in case of a conical base with a parabolic top with $h/R_0 = 35009$, $\cc_1 = -0.296$
and $\cc_2 = -1.116$. The numerically measured value of $\gamma_a$ is 6298 while
Eq.~\ref{eq:main} predicts 6294.

Eq.~\ref{eq:main} is valid for a single emitter rather than a family of emitters having a similar
base (such as conical). For a set of 10 emitters with a conical base and $h/R_a$ varying between
5000 to 200000, the best fitted value of $\cc_1 = -0.11657$ and $\cc_2 = -0.21655$. The average
error in prediction of $\gamma_a$ was found to be 2.02\%.
As an alternate formula for field enhancement,  Eq.~\ref{eq:fef} was fitted on the same
data set of conical emitters with respect to the parameter $\alpha$. The best fit was for
$\alpha = 0.88937$ and the average error in predicting $\gamma_a$ was 1.94\%.

\begin{figure}[htb]
%\vskip -2.1 cm
%\hskip -1.0cm
%\centering
\hspace*{-1.0cm}\includegraphics[width=0.5\textwidth]{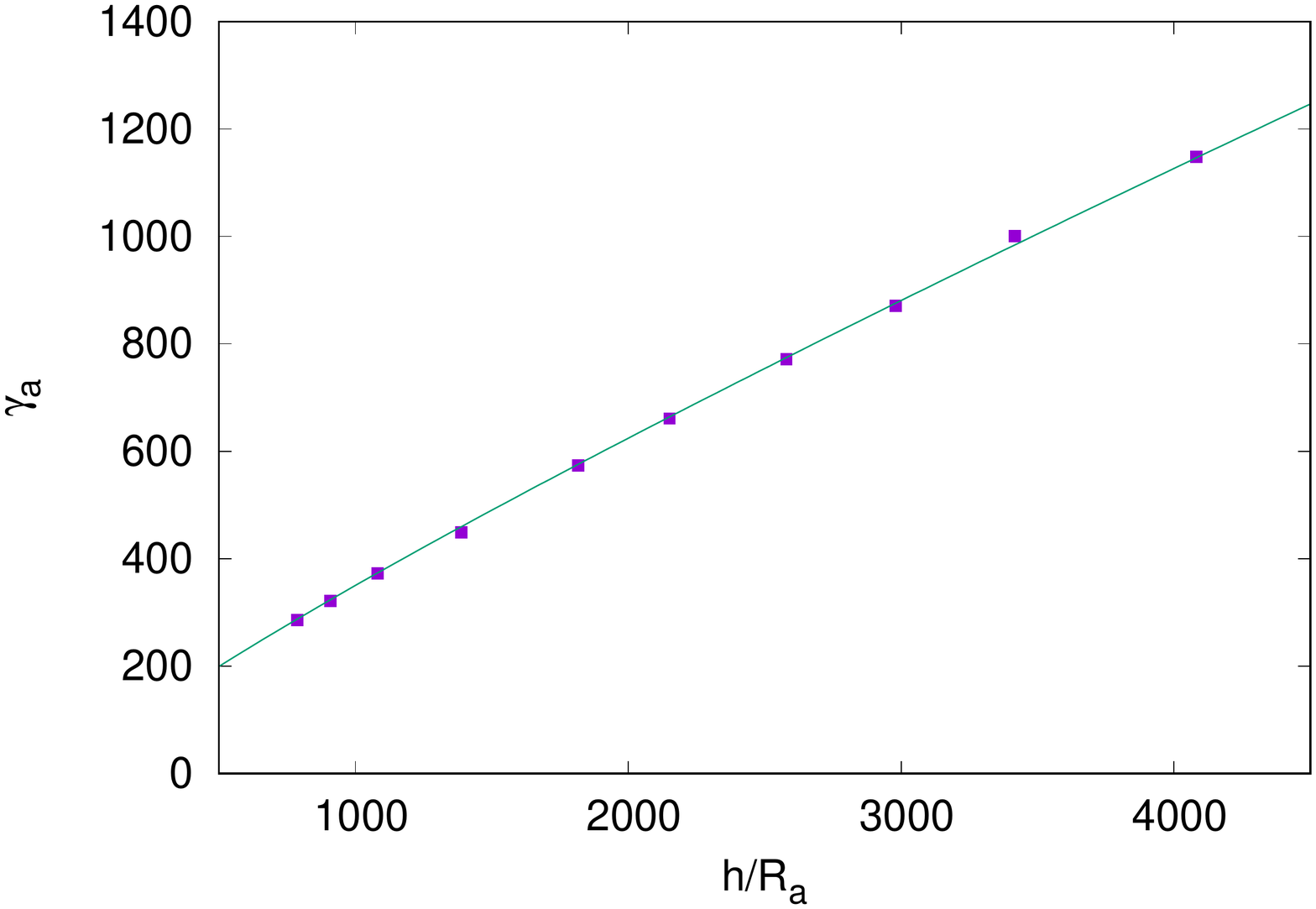}
\vskip -0.9 cm
\caption{The field enhancement factor for parabolic emitter tops with a cylindrical base (squares)
  with the best fit using Eq.~\ref{eq:fef} (continuous line).
}
\label{fig:angular}
\end{figure}

Similarly for a family of 10 parabolic emitter tops on a cylindrical base, the fitted
values using Eq.~\ref{eq:main} are $\cc_1 = 0.07611$ and $\cc_2 = 0.062489$.
On using Eq.~\ref{eq:fef}, the fitted value of the parameter is $\alpha = 2.5794$.
The average error in prediction is 0.73\% for Eq.~\ref{eq:main} and 0.76\% for Eq.~\ref{eq:fef}.

\section{Summary and Conclusions}

We have derived a formula for the field enhancement factor based on the line charge model.
It reduces to the result for ellipsoid that can be obtained by solving Laplace's
equation in prolate spheroidal co-ordinates. In general, it depends on parameters
that can be evaluated only when the line charge distribution is known numerically.
Interestingly, a set of optimally chosen parameters represents an entire family
of emitters and can be used with acceptable errors.

An alternate formula (Eq.~\ref{eq:fef}) that generalizes the known result for ellipsoid and hyperboloid
was also used to study a set of emitters with conical and cylindrical bases respectively. The
parameter $\alpha$ is known to be 0 for the hyperboloid where shielding of field lines
by the base is large. The conical base was found to have $\alpha \simeq 0.89$. The ellipsoid
is known to have $\alpha = 2$ while it was found to be $2.58$ for a cylindrical base. Thus,
$\alpha$ appears to be an indicator of shielding by the emitter base with a larger value
corresponding to lower shielding. We expect Eq.~\ref{eq:fef} to be of relevance even
in emitter arrays/clusters where electrostatic shielding is significant when the mean
spacing is smaller than the emitter height.

\section{Appendix: The Apex Radius of Curvature}

The apex radius of curvature $R_a$ can be expressed as
$R_a = (\partial V/\partial z)/(\partial^2 V/\partial \rho^2)$
evaluated at the apex. An expression for $\partial V/\partial z$ at the apex
has already been derived and the form in Eq.~\ref{eq:Vz} will be used. We shall
now arrive at a form for $\partial^2 V/\partial \rho^2$. On differentiating
Eq.~\ref{eq:pot} twice and evaluating at the apex,

\be
\frac{\partial^2 V}{\partial \rho^2} = -\frac{1}{4\pi\epsilon_0} \Big[ \int_0^L \frac{s f(s)}{(h-s)^3}~ds
  - \int_0^L \frac{s f(s)}{(h+s)^3}~ds \Big].
\ee

\noi
This can be expressed as

\be
\begin{split}
  \frac{\partial^2 V}{\partial \rho^2} = -\frac{1}{4\pi\epsilon_0} & \Big[ \int_0^L ds \Big\{ - \frac{f(s)}{(h-s)^2}   - \frac{f(s)}{(h+s)^2}~+ \\
    &  \frac{h f(s)}{(h-s)^3} + \frac{h f(s)}{(h+s)^3} \Big\} \Big]
\end{split}
\ee

which on integration by parts yields

\be
\begin{split}
  \frac{\partial^2 V}{\partial \rho^2} = \frac{1}{4\pi\epsilon_0} & \Big[ f(L) \frac{2L}{h^2 - L^2}(1 - \cc_0) \\
    & - \frac{2h^2 L f(L)}{(h^2 - L^2)^2} (1 - \cc_3) \Big]  \label{eq:app1}
\end{split}
\ee

\noi
where

\be
\cc_3 = \int_0^L \frac{f'(s)}{f(L)} \frac{s/(h^2 - s^2)^2}{L/(h^2 - L^2)^2}.
\ee

\noi
Using Bernstein's inequality again, $\cc_3 \sim (h^2 - L^2)^{1/2}$ which is
vanishingly small for sharp emitters. Further, the second term in Eq.~\ref{eq:app1} is dominant. Thus,

\be
\frac{\partial^2 V}{\partial \rho^2} \simeq -\frac{1}{4\pi\epsilon_0}  \Big[ \frac{2h^3 f(L)}{(h^2 - L^2)^2}  \Big] 
\ee

\noi
and

\be
\frac{\partial V}{\partial z} \simeq -\frac{1}{4\pi\epsilon_0} \frac{2 f(L) h^2}{h^2 - L^2}
\ee

\noi
so that $R_a \simeq \frac{h^2 - L^2}{h}$. Thus $R_0 \simeq R_a$.

\section{Acknowledgements} The author acknowledges useful discussions with Raghwendra Kumar, Gaurav Singh and Rajasree.

\vskip 0.05 in
%$\;$\\
%\section{References} 
\vskip -0.25 in
%\begin{references}


\begin{thebibliography}{99}
  
\bibitem{FN} R.~H.~Fowler and L.~Nordheim, Proc. R. Soc. A 119, 173 (1928).
\bibitem{murphy} E.~L.~Murphy and R.~H.~Good, Phys. Rev. 102, 1464 (1956).
\bibitem{forbes} R.~G.~Forbes, App. Phys. Lett. 89, 113122 (2006). 
\bibitem{forbes_deane} R.~G.~Forbes and J.~H.~B.~Deane, Proc. Roy. Soc. A 463, 2907 (2007).
\bibitem{fursey} G.~Fursey, {\it Field emission in vacuum microelectronics},  Kluwer/Plenum, New York, (2005)
\bibitem{jensen_ency} K.~L.~Jensen, {\it Field emission - fundamental theory to usage},
  Wiley Encycl. Electr. Electron. Eng. (2014).
\bibitem{liang} Shi-Dong Liang, {\it Quantum Tunneling and Field Electron Emission Theories}, World Scientific Publishing Co. Pte. Ltd., Singapore (2014).
  \bibitem{egorov_springer} N.~Egorov and E.~Sheshin, {\it Field Emission Electronics}, Springer Series in Advanced Microelectronics, Vol. 60 (2017).
\bibitem{db2018a} D.~Biswas, https://arxiv.org/abs/1801.09508.
\bibitem{forbes2003} R.~G.~Forbes, C.J.~Edgcombe and U.~Valdr\`{e}, Ultramicroscopy 95, 57 (2003).
\bibitem{podenok} S. Podenok, M. Sveningsson, K. Hansen and E.E.B. Campbell, Nano 1, 87 (2006).
\bibitem{read} F.~H.~Read and N.~J.~Bowring, Nucl. Ins. Meth. Phys. Res. 519, 305 (2004).
\bibitem{kokkorakis} G.~C.~Kokkorakis, A.~Modinos, J.~P.~Xanthakis, J. Appl. Phys. 91, 4580 (2002).
\bibitem{bonard} J-M. Bonard, K.~A.~Dean, B.~F.~Coll and C.~Klinke, Phys. Rev. Lett. 89, 197602 (2002).
\bibitem{zhbanov} A. I. Zhbanov, E. G. Pogorelov, Y.-C. Chang, and Y.-G. Lee, J.~Appl.~Phys. 110, 114311 (2011).
\bibitem{forbes2016} R.~Forbes, J.~App.~Phys. 120, 054302 (2016).
\bibitem{jap2016} D.~Biswas, G.~Singh and R.~Kumar, J.~App.~Phys. 120, 124307 (2016).
\bibitem{db_ultram} D.~Biswas, G.~Singh, S.~G.~Sarkar and R.~Kumar, Ultramicroscopy 185, 1 (2018).
\bibitem{ineq} See for example, R.~Whitley, J.~Mathematical Analysis and Applications, 105, 502 (1985).

  
  \end{thebibliography}
\end{document}